\documentclass[aps,twocolumn,showpacs]{revtex4}

\usepackage{epsfig}

\begin{document}

\title{Brownian forces in sheared granular matter}

\author{A. Baldassarri$^1$, F. Dalton$^2$, A. Petri$^2$,
  S. Zapperi$^{3}$, G. Pontuale$^2$, and L. Pietronero$^{1,2}$}

\affiliation{$^1$Dipartimento di Fisica,
Universit\`a "La Sapienza", P.le A. Moro 2, 00185 Roma, Italy\\
$^2$ CNR, Istituto dei Sistemi Complessi, sede di Tor Vergata, Via del Fosso del Cavaliere 100, 00133 Roma, Italy.\\
$^3$ CNR-INFM Universit\`a "La Sapienza", P.le A. Moro 2, 00185 Roma, Italy}

\begin{abstract}
We present
results from a series of experiments on a granular medium sheared
in a Couette geometry and show that their statistical properties
can be computed in a quantitative way from the assumption that the
resultant from the set of forces acting in the system performs a
Brownian motion. The same assumption has been utilised, with
success, to describe other phenomena, such as the Barkhausen
effect in ferromagnets, and so the scheme suggests itself as a
more general description of a wider class of driven
instabilities.
\end{abstract}

\maketitle

The shear response of granular media has been investigated
for more than a century by different scientific communities
\cite{REY-85}, ranging from soil mechanics~\cite{TAY-48} and
earth sciences \cite{MAR-98,SEG-95} to physics \cite{BAG-66,NAS-97}.
This fact is reflected by the variety of experimental settings and
theoretical methodologies applied to the problem, such as rate and
state friction laws \cite{SEG-95,LAC-00} or direct particle
simulations~\cite{THO-91,TIL-95}.  Under a slow loading rate, the
shear response of granular media typically displays large
fluctuations~\cite{MIL-96,DAL-02}, due to the rearrangements
of the internal force chains~\cite{ALB-99,GEN-05} along which stress
propagates in granular materials.

Motivated by the practical concern of obtaining a mechanical
description to be used in applications,  most of the past
theoretical activity has been devoted to the analysis of averaged
properties, in the search for constitutive macroscopic laws. A
detailed analysis of slip statistics is, however, of primary
importance if one wishes to understand the mechanisms which are at
the base of the wide fluctuations observed. They are the
distinguishing feature of "crackling noise", which denotes an
intermittent activity with widely fluctuating amplitudes
\cite{SET-01}.  Crackling noise is observed in different contexts
such as magnetic materials \cite{BAR-19,Bertotti}, ferroelectrics,
type II superconductors \cite{FIE-95}, fracture
\cite{PET-94,GAR-97} and plasticity \cite{ANA-99,MIG-01} to name
just a few. In addition it shares analogies with earthquake
phenomenology, where the statistics typically displays wide
fluctuations and recorded sizes of seismic events span several
decades \cite{GUT-44,MAI-96}. Common aspects observed in such
apparently different systems have suggested the existence of a
deeper correspondence in the underlying physical properties
\cite{SET-01}, and one may hope that any general insight or
understanding gained in these phenomena may guide research in
other fields. Among the schemes proposed in this context, perhaps
the most successful are the scaling-renormalization group
\cite{SET-01} and Self-Organized Criticality~\cite{JEN-98}. Within
these frameworks many phenomena have been qualitatively explained
in terms of a set of logical, coherent and simple rules, but it is
generally difficult to adapt them to specific situations and
obtain quantitative predictions.  In this letter, starting from a
laboratory experiment on sheared granular matter we propose a
model which quantitatively reproduces the observations and, at the
same time, is based on a simple but intuitive hypothesis which
sheds light on the similarities between different
phenomena.

\begin{figure}[t]
\centerline{
\includegraphics[width=7cm]{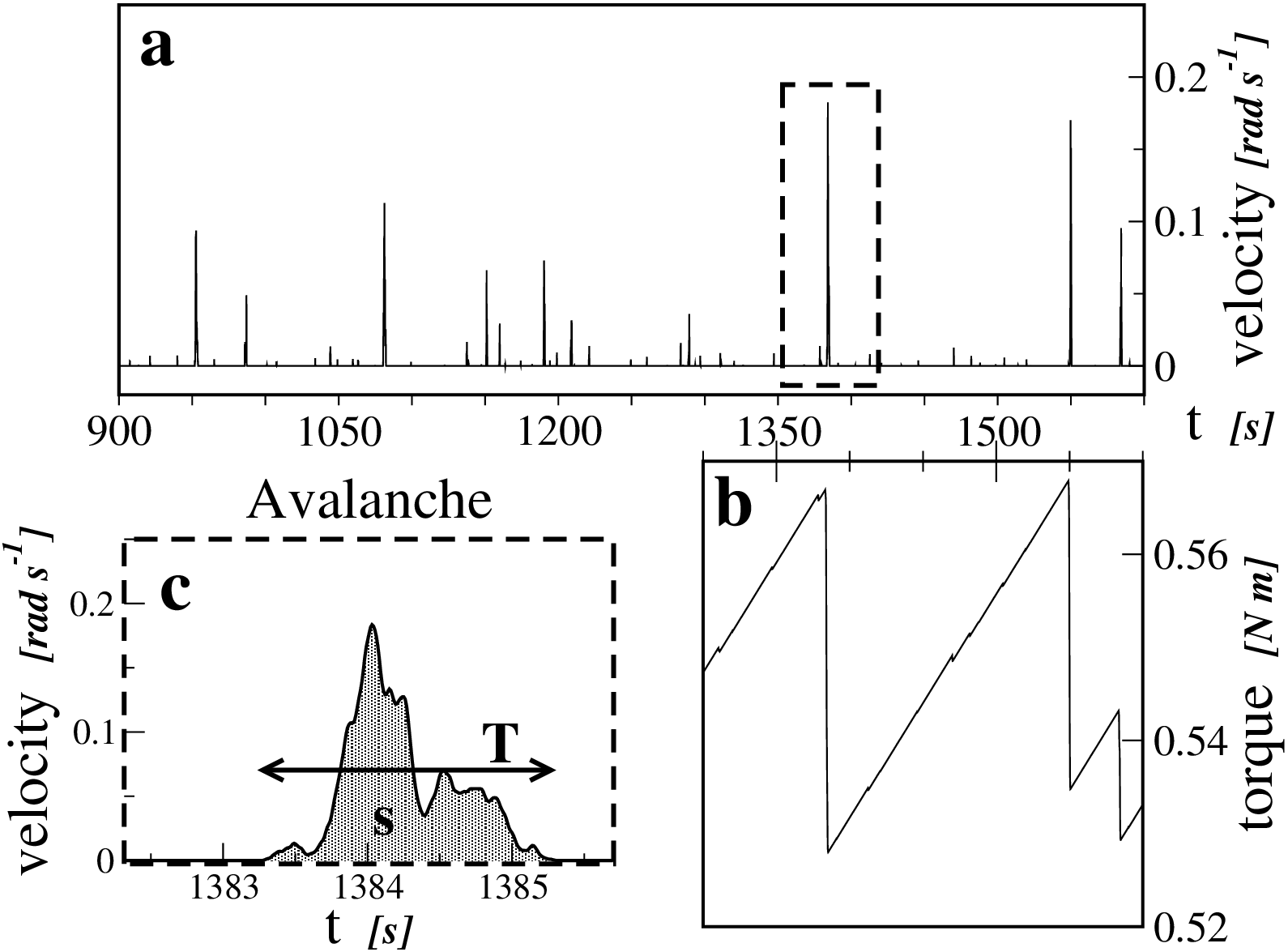}
        }
\caption{The signals measured in experiments: a) instantaneous velocity
  of the disk; b) instantaneous torque exerted by the spring;
  c) definition of duration and size of a slip event.}
\label{fig:0}
\end{figure}

Our experimental set-up consists of a Couette cell made of a circular channel
containing mono-disperse $2$mm glass beads.  An annular plate is driven over
the top surface of the channel by a motor through a torsion spring with
stiffness $k$. In most of the experiments we employ a spring of stiffness
$k=0.36$N, but we also consider different values (i.e. $k=0.12$N and $0.9$N).
To ensure a granular shearing plane, the annular top plate has a layer of
beads glued to its lower surface, though, to avoid individual grains jamming
the system at the boundary, this layer of beads does not extend the full width
of the channel. The system is initialized before each experiment by pouring
the medium into the channel, then the system is run at a slow velocity $V$
($\simeq 0.01$ rad/s) for a long time, in order to approach a stationary
state.  We have observed that the mean torque approaches a steady-state after
the order of $100$ revolutions.  Our measurements from the device consist of the
angular position $\theta$ of the top plate and the deflection of the torsion
spring, sampled at 1 kHz with an error $\simeq 10^{-4}$ rad. From this, with
knowledge of the torsion spring constant and the moments of inertia of the
plate, we obtain angular velocity and acceleration, instantaneous torque and
work done by the spring. Varying the driving angular
velocity $V$ in the range $10^{-4}$-$10^{-1}$ rad/s, the system displays a the
stick-slip regime: the disk is still as the spring winds up and the torque on
the plate increases until the plate slips. The equation of motion of the disk reads:
\begin{equation}
 I \ddot {\theta} = k( Vt -\theta) - F , \label{eq:model}
\end{equation}
where $I$ is the moment of inertia of the system, $k$ is the stiffness of the
torsional spring, and $V$ is the angular velocity of the driving
motor. The first term on the righthand side is the force
exerted by the spring. The motion of the plate is counteracted by
the ''friction" $F$ exerted by the medium. Is this last term that encompass the
whole complexity of the physical stick-slip phenomena.

\begin{figure}[t]
\centerline{
\includegraphics[width=7cm]{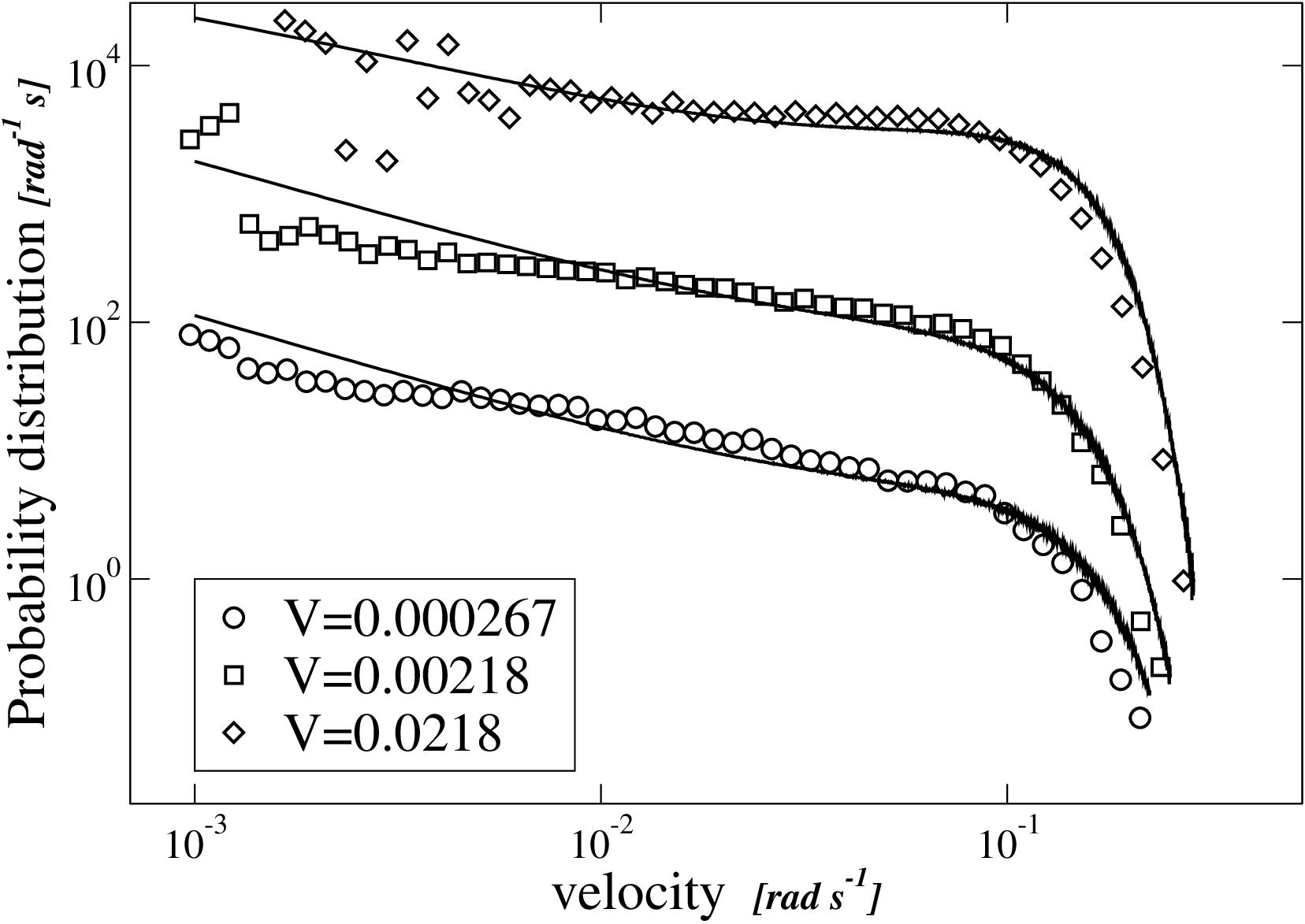}
        }
\caption{The experimentally measured distribution of disk velocities for different values of
the driving angular velocity $V$. The lines are obtained from the numerical integration of
Eqs.~(\ref{eq:model}-\ref{eq:dF}) using parameters value extracted from the
experiments. The distributions for different drives are vertically shifted for clarity.}
\label{fig:1}
\end{figure}

\begin{figure}[t]
\centerline{
\includegraphics[width=7.0cm]{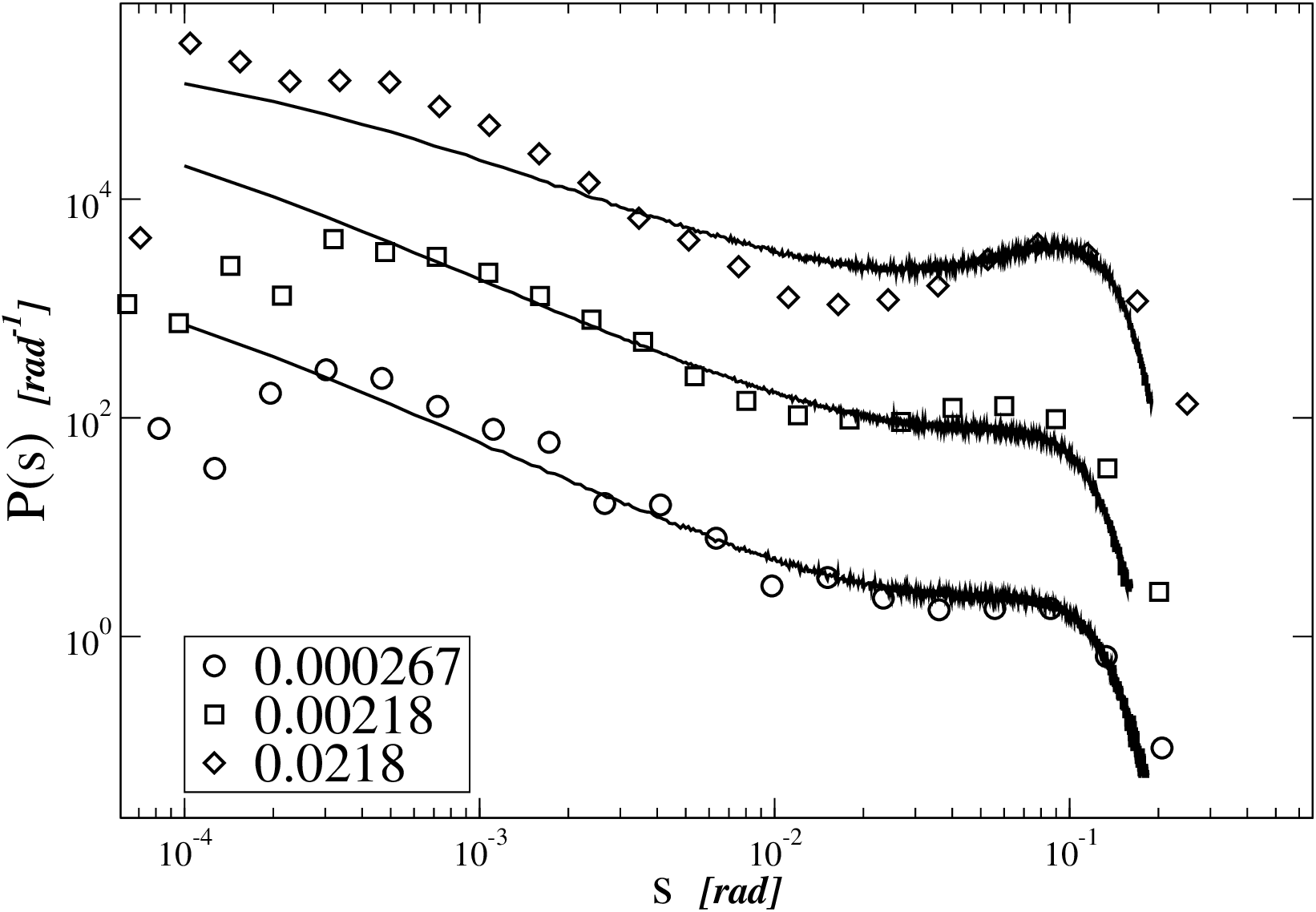}
}
\caption{Slip sizes probability distribution, for different values of the
  driving angular velocity. The lines are the results from the numerical integration of
Eqs.~(\protect\ref{eq:model}-\protect\ref{eq:dF}). The distributions for different drives are vertically shifted for clarity.}
\label{fig:2a}
\end{figure}

\begin{figure}[t]
\centerline{
\includegraphics[width=7.0cm]{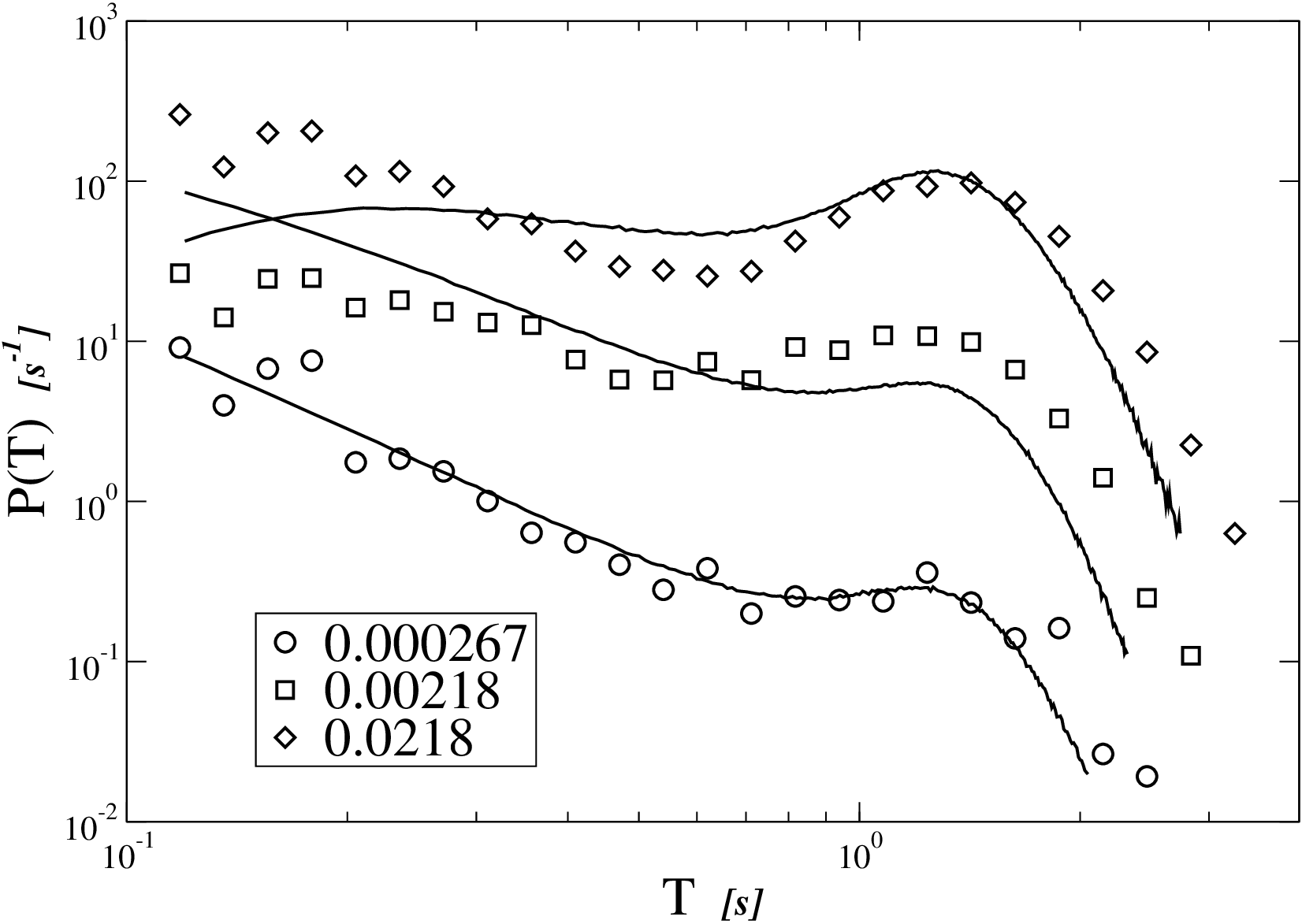}
}
\caption{Slip durations probability distribution, for different values of the
  driving angular velocity. The lines are the results from the numerical integration of
Eqs.~(\protect\ref{eq:model}-\protect\ref{eq:dF}). The distributions for different drives are vertically shifted for clarity.}
\label{fig:2b}
\end{figure}

The typical instantaneous velocity signal displays a very erratic
behavior with pulses of widely fluctuating magnitudes (see
Fig.~\ref{fig:0}a), and its probability density distribution 
spans more than two decades. The presence of power law statistics in the
stick-slip phase of a sheared granular was first reported some
years ago in slightly different experiments \cite{DAL-02}
The present experiments show that the general shape of the distributions
depends on the shear rate, at odds with the cutoff position (see Fig.~\ref{fig:1}).
One could attempt to fit the data in Fig.~\ref{fig:1} with a power law
to obtain an exponent, but the limited scaling regime would not yield a reliable 
result. In order to get a more complete characterization of the fluctuations
and their temporal scales, we analyze duration $T$ and a size $S$
associated to each pulse, as is commonly done in crackling noise
experiments \cite{SET-01}; in this case $S$ is equal to the angular slip of the
plate during the pulse (see Fig.~\ref{fig:0}c). The distributions
of $T$ and $S$, measured for different values of the driving
angular velocity $V$, are reported in Fig.~\ref{fig:2a} and Fig.~\ref{fig:2b}. They
display a complex shape, in which an initial decay is followed by a
characteristic peak at larger scales, indicading broad fluctuations
over many scales. Note that the position of the peaks does not depend 
on the driving velocity $V$.

This behavior is entirely contained within $F$ of
Eq.~(\ref{eq:model}), the instantaneous frictional torque. Because
of the disordered arrangement of grains, forces propagate within the
medium via a complicated network in which both elastic and
frictional interaction play fundamental roles~\cite{GOLD-05}.
Moreover the network is constantly evolving in a pseudo-plastic way
\cite{MAJ-05}, and so a detailed force description represents a
formidable problem with a very large number of interdependent
degrees of freedom.  The friction term $F$ is, in principle, a random function which may
depend on a variety of state variables~\cite{MAR-98}. The minimal formulation
retains only the dependence on
the position and velocity of the disk, i.e. we assume that
$F=F(\theta,\dot\theta)$. Note that the dependence on the time arises only from
$\theta$ and $\dot\theta$, since when the disk is stuck the force does not change.
As an additional assumption, we split $F$ into two
additive terms: In the first of them, the dependence on velocity is confined
to a deterministic law, analogous to the viscous reaction of a fluid, while
the second term is independent of the velocity and introduces the fluctuating
part: $F(\theta,\dot{\theta})=\bar{F}(\dot{\theta})+F_f(\theta).$

\begin{figure}[ht]
\centerline{
\includegraphics[width=7cm]{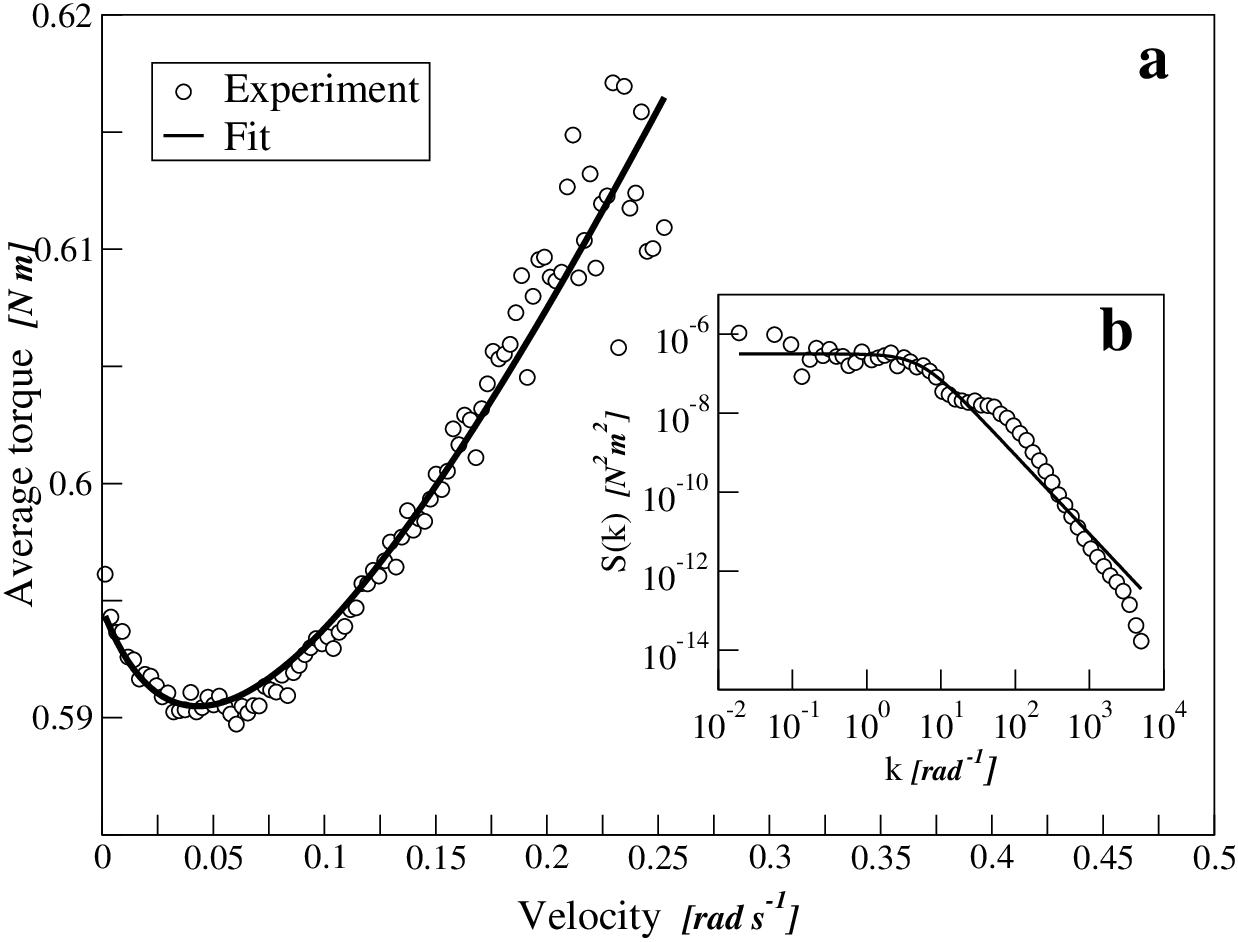}
        }
\caption{Frictional torque: a) average velocity dependence; b)
power spectrum of the random frictional torque.}\label{fig:4}
\end{figure}

In Fig.~\ref{fig:4}a we report the value of the friction torque as
a function of the instantaneous angular velocity of the plate.
Once fluctuations have been averaged out, a clear velocity
dependence is seen, with strain-rate weakening at low velocity
followed by strengthening at high velocity. These features, not
present in the viscous reaction of a Newtonian fluid, are
responsible for the stick-slip instability. An adequate fit of the
experimental data is provided by a rate dependent friction law of
the type

\begin{equation}
\bar{F}(\dot\theta)=F_0 + \gamma (\dot\theta -2v_0 \ln(1+\dot\theta/v_0)), 
\label{eq:Fv}
\end{equation}

where $F_0$ is the average static friction torque, $v_0$ corresponds
to the minimum in the average torque and $\gamma$ is the high
velocity damping (see Fig.~\ref{fig:4}a).

It is interesting to note that an equation similar to Eq.~(\ref{eq:Fv}) 
is used to describe the friction law observed in dry solid-on-solid
friction~\cite{HES-94} (steady sliding regime).
Previous work have shown that the friction force from granular
materials is not a single valued function of the velocity during a
single slip event~\cite{NAS-97}. Our analysis does not contradict
this observation, since the result displayed in 
Fig.~\ref{fig:4}a describes the average over different slip
events.

The fluctuating part of the friction torque is, as discussed above,
a direct consequence of the disordered structure of the force chain
network present in the granular medium. As the disk slips by a small
angle $\delta\theta$, one expects that the friction torque is
increased or decreased by a small random amount $\delta F_f$,
reflecting  a limited rearrangement in the granular structure. 
On the other hand,
under subsequent or large slips, rearrangement of grains will be
more complete, and the random torque will eventually decorrelate.
The simplest mechanism to account for these qualitative features is
a confined {\em Brownian} process:
\begin{equation}\label{eq:dF}
\frac{dF_f}{d\theta}=\eta(\theta)-a \theta,
\end{equation}
where $\eta$ is an uncorrelated noise term extracted from a Gaussian distribution, i.e.
$\langle \eta(\theta)\eta(\theta') \rangle = D
\delta(\theta-\theta')$, where $D$ is the noise variance.
It is important to note that the independent variable here 
is the angular position $\theta$ instead
of the time $t$, as always the case for quenched athermal disordered systems.

The parameter $a$ determines the inverse
correlation length and can be estimated by the power spectrum
$S(\omega)$ of the process. In fact this latter is related to the
correlation function by the Wiener-Kintchine theorem. For the process
above the power spectrum is given by
\begin{equation}
S(k)=\langle \left|\int d\theta  F_f(\theta) \exp(-i \theta k)\right|^2 \rangle =
\frac{2 D}{a^2 + k^2}.
\label{eq:CF_f}
\end{equation}
The power spectrum measured in the experiment is shown in
Fig.~\ref{fig:4}b. It is seen that despite the simplicity of
Eq.(\ref{eq:dF}), the related Lorentzian spectrum of
Eq.(\ref{eq:CF_f}) compares quite well with the actual spectrum and
allows estimation of the inverse correlation length $a$. Once the
parameters $a$, $D$, $F_0$, $\gamma$ and $v_0$ have been extracted
from the data, the equation of motion (\ref{eq:model}) is integrated
numerically, and a stochastic velocity signal is obtained, to be
compared with the experiments. It should be emphasized that the the fitted 
parameters display quite stable values upon repeating the experiments and varying
the driving rate. Typical values at low $V$ are $a \approx 15$ rad$^{-1}$, 
$D \approx 0.01$ (N$^2$/rad),
$F_0 \approx 0.55$ Nm, $\gamma \approx 0.20$ N m s /rad, $v_0 \approx  0.03$ rad/s. At high velocities
($V>0.1$ rad/s) we observe a significant variation of  these values,
indicating the onset of the fluidization transition \cite{DAL-05}. 

The model reproduces the complex phenomenology of the granular dynamics, despite being
based on few mean-value parameters whose values are directly measured
from the system under study. The only unknown parameter here is the moment of
inertia of the system $I$.  In fact, considering $I$ as the bare moment of
inertia of the disk $I_0$, one neglects the possibility that a number of grains
could move together with the plate, increasing the actual moment of inertia of
the system. On the contrary, this behavior is expected for the granular system, due to
formation of shear bands~\cite{LAC-02}. Accordingly, a best quantitative agreement 
between model and experiments is obtained considering $I=1.5 I_0$, where
$I_0=0.02$ Kg m$^2$, corresponding to few granular layers moving together with
the disk.

Examples of the distributions resulting from the model are compared to experimental data in Fig.~\ref{fig:1},
Fig.~\ref{fig:2a} and Fig.~\ref{fig:2b}. We notice that the best agreement is 
found at low driving. Possible reasons for the discrepancies observed at high drives
are related to the presence of the fluidization transition and the intrinsic approximations
of our model. In particular, the power spectrum is not exactly Lorentzian and the model does 
not account for dilatancy effects and complex shear band dynamics. Nevertheless, the simplicity of 
the proposed approach is quite attractive and suggests the tantalizing idea
that phenomena observed in different contexts could be effectively
described  by similar general processes.

\begin{table}[t]
\begin{center}
\begin{tabular}{|lc|r|}
\hline
slip & $\theta$&   domain wall jump\\
shear rate & $kV$ &  magnetic field rate\\
spring constant & $k$ & demagnetizing factor\\
frictional damping & $\gamma$ & eddy current damping\\
random friction & $F_f$ & pinning forces \\
moment of inertia & $I$ & domain wall mass\\
\hline
\end{tabular}
\caption{Table of correspondences between physical quantities for
the shear response of a granular medium and the domain wall motion
in a ferromagnetic material.}
\label{fig:6}
\end{center}
\end{table}

This idea is supported by the observation that a similar process
emerges in a very different context from the present one,
specifically that of ferromagnetic materials.  
The equation of motion we adopt to describe granular media is similar to the one
employed for domain wall motion~\cite{ALE-90}.  The latter reproduces
the statistical properties of the Barkhausen noise
recorded during hysteresis in soft magnetic materials. In particular,
the granular equation reduces to the domain wall model in the limits
$V>> v_0$ and $\gamma >> I \cdot V$, as is appropriate for metallic
ferromagnets where eddy current damping cancels inertial effects. The
random structure of the friction force in the case of domain walls is
due to the inhomogeneities present in the material and the correlation
properties are believed to arise from a summation of the pinning
forces along the domain wall \cite{ZAP-98}, while here they are due to
the formation and destruction of force chains. A summary of the
mapping between granular matter and magnetic domain walls is reported
in Table~\ref{fig:6}.

The domain wall model is exactly solvable and predicts a power law dependence
for the distribution of velocities, slip sizes and durations \cite{ALE-90};
the corresponding exponents decrease linearly with the driving rate.  The
power laws are terminated by an exponential cutoff, inversely proportional to
the demagnetizing factor~\cite{ZAP-98}.

Our granular model presents an additional feature: the distribution tails
display a peak that masks the power law decay. 
This is due to the combined effect of inertia and the shear rate
weakening frictional torque. This idea is confirmed by the fact that the peak
in the duration distribution scales as $T_p \sim \sqrt{I/k}$. This behavior is
simply recovered considering the deterministic motion obtained by suppressing
friction fluctuations in Eq.~(\ref{eq:model}), and it has been confirmed in
our experiments.  Future experiments, designed to reduce inertial
effects and shear rate weakening, could allow the observation of the power law
decays, whose exponents are predicted by our model.

The results presented here, supported by the results for the
Barkhausen effect in ferromagnets~\cite{ALE-90}, indicate that the
Brownian scheme (eq.~\ref{eq:dF}) for the description of random forces evolution in
dissipative systems driven by an external force may constitute an
effective approach of general validity.
Similar analysis of different
systems exhibiting "crackling noise"~\cite{SET-01} will constitute a
proving ground for these ideas.

\end{document}